\documentclass[conference,twocolumn]{IEEEtran}
\usepackage{amsmath}
\usepackage{amssymb}
\usepackage[dvipdfmx]{graphicx}
\usepackage{ascmac}
\ifCLASSINFOpdf
\else
\fi
\begin{document}

\sloppy

%% Paper Title
%% You can use linebreaks \\ within to get better formatting as
%% desired. 
\title{Variable-Length Resolvability for Mixed Sources and its Application to Variable-Length Source Coding }

%% Author names and affiliations:
%%
\author{
  \IEEEauthorblockN{{\normalsize Hideki Yagi~~~}}
  \IEEEauthorblockA{
Dept.\ of Computer \& Network Engineering\\
University of Electro-Communications \\
Tokyo, Japan\\
{\normalsize Email: h.yagi@uec.ac.jp}} 
  \and
  \IEEEauthorblockN{Te Sun Han}
  \IEEEauthorblockA{National Institute of Information and \\
   Communications Technology (NICT)\\
    Tokyo, Japan\\
    Email: han@is.uec.ac.jp}
}

%% To balance the two columns, you should reduce the text-height of
%% the last page using the following command:
%%%%%%%%%%%%%%%%%%%%%%%%%%%%%%%%%%%%%%%%%%%%%%%%%%%%%%%%%%%%%%%%%%%%%
%\addtolength{\textheight}{-9.35cm}
%%%%%%%%%%%%%%%%%%%%%%%%%%%%%%%%%%%%%%%%%%%%%%%%%%%%%%%%%%%%%%%%%%%%%
%% with an appropriate value. This command must be place on the second
%% last page, i.e., for a one-page abstract here, for a two-page
%% abstract right after the \maketitle command.
%\newcommand{\limsup}{\textrm{limsup}}		% This is my own macro !!!
\newtheorem{e_theo}{Theorem}
\newtheorem{e_defin}{Definition}
\newtheorem{e_lem}{Lemma}
\newtheorem{e_prop}{Proposition}
\newtheorem{e_coro}{Corollary}
\newtheorem{e_rema}{Remark}
\newtheorem{e_exam}{Example}

\newcommand{\SrAlphabet}{{\mathcal{U}}}
\newcommand{\CdAlphabet}{{\mathcal{X}}}
\newcommand{\CoinSr}{{U^n}}
\newcommand{\TargetSr}{{X^n}}
\newcommand{\ApproxSr}{{\tilde{X}^n}}
\newcommand{\CoinSeq}{{\vect{u}}}
\newcommand{\TargetSeq}{{\vect{x}}}
\newcommand{\E}{\mathbb{E}}	
\newcommand{\V}{\mathsf{V}}	
\def\vect#1{\boldsymbol{#1}}
\def\QED{\hfill$\Box$}

%% Create the title:
\maketitle

%% Abstract: 
%%
\begin{abstract}
In the problem of variable-length $\delta$-channel resolvability, the channel output is approximated by encoding a variable-length uniform random number under the constraint that the variational distance between the target and approximated distributions should be within a given constant $\delta$ asymptotically.
In this paper, we assume that the given channel input is a mixed source whose components may be general sources. 
To analyze the minimum achievable length rate of the uniform random number, called the $\delta$-resolvability, we introduce a variant problem of the variable-length $\delta$-channel resolvability.
A general formula for the $\delta$-resolvability in this variant problem is established for a general channel.
When the channel is an identity mapping, it is shown that the $\delta$-resolvability in the original and variant problems coincide. This relation leads to a direct derivation of a single-letter formula for the $\delta$-resolvability when the given source is a mixed memoryless source.
We extend the result to the second-order case. As a byproduct, we obtain the first-order and second-order formulas for fixed-to-variable length source coding allowing error probability up to $\delta$. 
\end{abstract}

% no keywords

\renewcommand\thefootnote{}
\footnotetext{This research is supported by JSPS KAKENHI Grant Numbers JP16K06340 and 	JP17K00020.}
\renewcommand\thefootnote{\arabic{footnote}}

\IEEEpeerreviewmaketitle

%===========================================
%============  Section VI  ================
%===========================================
\section{Introduction} \label{sec:introduction}

In the problem of \emph{variable-length} $\delta$-channel resolvability, the channel output is approximated by encoding a variable-length uniform random number under the constraint that the distance (e.g.\ variational distance) between the target and approximated distribution should be within a given constant $\delta$ asymptotically.
This problem, introduced by Yagi and Han \cite{Yagi-Han2017}, is a generalized form of the \emph{fixed-length} $\delta$-channel resolvability \cite{Han-Verdu93,Hayashi2006} in which the fixed-length uniform random number is used as a coin distribution.
The minimum achievable length rate of the uniform random number, referred to as the $\delta$-\emph{resolvability}, is the subject of analysis.
In \cite{Yagi-Han2017}, a general formula for  the $\delta$-resolvability has been established for any given source and channel.
Recently, a single-letter formula for the $\delta$-resolvability has been given in \cite{Yagi-Han2017sita} when the source and the channel are stationary and memoryless.
An interesting  next step may be a \emph{mixed memoryless sources} and/or a \emph{mixed memoryless channel} \cite{Han2003}, which are stationary but non-ergodic stochastic processes.

In this paper, we assume that the given channel input is a mixed source with components which may be general sources. 
To establish a general formula of the $\delta$-resolvability for a general channel, we introduce a variant problem of the variable-length $\delta$-channel resolvability.
When the channel is an identity mapping, it is shown that the $\delta$-resolvability in the original and variant problems coincide.
This relationship is of use to derive a single-letter formula for the $\delta$-resolvability when the given source is a mixed memoryless source.
We also extend the result to the second-order case.
It is known that the $\delta$-resolvability coincides with the minimum achievable coding rate of the weak fixed-to-variable length (FV) source coding allowing  error probability up to $\delta$.
As a byproduct, we obtain the first-order and second-order formulas for this minimum achievable coding rate. 

%===========================================
%============  Section VI  ================
%===========================================
\section{Problem of Variable-Length Channel Resolvability} \label{sec:channel_resolvability}

In this section, we review the problem of channel resolvability in the variable-length setting.

Let $\mathcal{X}$ and $\mathcal{Y}$ be finite or countably infinite alphabets.
Let $\vect{W} = \{W^n \}_{n=1}^\infty$ be a general  channel, where $W^n : \mathcal{X}^n \rightarrow \mathcal{Y}^n$ denotes a stochastic mapping.
We denote by $\vect{Y} = \{ Y^n \}_{n=1}^\infty$ the output process {via} $\vect{W}$ due to the input process $\vect{X}= \{ X^n \}_{n=1}^\infty$, {where $X^n$ and  $Y^n$ take values in $\mathcal{X}^n$ and $\mathcal{Y}^n$, respectively}. 
{The probability distributions of $X^n$ and $Y^n$ are denoted by} $P_{X^n}$ and $P_{Y^n}$, respectively, and these symbols are used interchangeably.

\medskip
Consider the problem of variable-length channel resolvability.
Let $\SrAlphabet^*$ denote the set of {all}  sequences $\CoinSeq \in \SrAlphabet^m$ {over} $m = 0, 1, 2, \cdots$, where $\SrAlphabet^0 = \{ \lambda \}$ ($\lambda$ is the null string). 
Let $L_n$ denote a random variable which takes {values} in $\{0,1,2, \ldots\}$.
We define {the} \emph{variable-length uniform random number} $U^{(L_n)}$ {so} that
$U^{(m)}$ is uniformly distributed over $\SrAlphabet^m$ given $L_n = m$.
{In other words, for $\CoinSeq \in \SrAlphabet^m$}
\begin{align}
P_{U^{(L_n)}}(\CoinSeq, m) &:= \Pr\{ U^{(L_n)} = \CoinSeq , L_n = m \} \nonumber \\
&=\frac{\Pr\{ L_n = m \}}{K^m}, 
\end{align}
where $K = |\SrAlphabet|$.
It should be noticed that {variable-length sequences} $\CoinSeq \in \SrAlphabet^m$ {are} generated with joint probability $P_{U^{(L_n)}}(\CoinSeq, m)$.
Consider the problem of approximating the target output distribution $P_{Y^n}$ {via} $W^n$ due to {$X^n$} by using another input $\tilde{X}^n=\varphi_n(U^{(L_n)})$ with a deterministic mapping (encoder) $\varphi_n :  \mathcal{U}^*  \rightarrow \mathcal{X}^n$. 
Let $d(P_{Y^n}, P_{\tilde{Y}^n}) := \frac{1}{2} \sum_{\vect{y}} |P_{Y^n} (\vect{y}) -  P_{\tilde{Y}^n}(\vect{y})| $ be the \emph{variational distance} between $P_{Y^n}$ and $P_{\tilde{Y}^n}$.

\medskip
\begin{e_defin} \label{def:VL_achievable_rate}
{\rm
Let $\delta \in [0, 1)$ be fixed arbitrarily.
A resolution rate $R \ge 0$ is said to be \emph{$\delta$-variable-length achievable} {or simply $\mathrm{v}(\delta)$-achievable} for $\vect{X}$ {(under the variational distance)} if there exists a variable-length uniform random number $U^{(L_n)}$ and a deterministic mapping $\varphi_n : \mathcal{U}^* \rightarrow \mathcal{X}^n$ satisfying
\begin{align}
 \limsup_{n \rightarrow \infty} \frac{1}{n} \E [L_n] &\le R, \label{eq:VL_ch_rate_cond}  \\ 
 \limsup_{n \rightarrow \infty} d(P_{Y^n}, P_{\tilde{Y}^n}) &\le \delta, \label{eq:VL_ch_variational_dist_cond}   
\end{align}
where $\E[\cdot]$ denotes the expected value and $\tilde{Y}^n $ denotes the output via $W^n$ due to the input $\tilde{X}^n = \varphi_n(U^{(L_n)})$.
The infimum of all {$\mathrm{v}(\delta)$-achievable} rates for $\vect{X}$:
\begin{align}
S_{\rm v} (\delta | \vect{X}, \vect{W}):= \inf \{ R : ~ R ~\mbox{is {$\mathrm{v}(\delta)$-achievable} for}~ \vect{X}\} \label{eq:VL_ch_opt_rate}
\end{align}
is called the \emph{$\delta$-variable-length channel resolvability} {or simply $\mathrm{v}(\delta)$-channel resolvability} for $\vect{X}$.}
\QED
\end{e_defin}

\medskip
When the channel $W^n$ is an identity mapping, the addressed problem reduces to {that of} \emph{source resolvability}.
\medskip
\begin{e_defin} \label{def:VL_achievable_rate2}
{\rm
Assume that the channel  $W^n$ is an identity mapping.
The infimum of all {$\mathrm{v}(\delta)$-achievable} rates for $\vect{X}$:
\begin{align}
S_{\rm v} (\delta | \vect{X}):= \inf \{ R : ~ R ~\mbox{is {$\mathrm{v}(\delta)$-achievable} for}~ \vect{X}\} \label{eq:VL_ch_opt_rate2}
\end{align}
is called the \emph{$\delta$-variable-length source resolvability} {or simply $\mathrm{v}(\delta)$-source resolvability} for $\vect{X}$.}
\QED
\end{e_defin}

\medskip
Let $\mathcal{P}(\CdAlphabet^n)$ denote the set of all probability distributions on $\CdAlphabet^n$.
For $\delta \in [0, 1]$, defining the \emph{$\delta$-ball} using the variational distance as
\begin{align}
B_{\delta}(\TargetSr) = \left\{ P_{V^n} \in \mathcal{P}(\CdAlphabet^n) : d(P_{\TargetSr}, P_{V^n}) \le \delta \right\},
\end{align}
we introduce the \emph{smooth entropy}:
\begin{align}
H_{[\delta]} (\TargetSr) &:=  \inf_{ P_{V^n} \in B_{\delta}(\TargetSr) }  H(V^n), \label{eq:smooth_reny_entropy}
\end{align}
where $H(V^n)$ denotes the Shannon entropy of $P_{V^n}$.
The $H_{[\delta]} (\TargetSr)$ is a nonincreasing {monotone} function of $\delta$. 
Based on this quantity {for} a general source ${\vect{X}} = \{ \TargetSr\}_{n =1}^\infty$, we define 
\begin{align}
H_{[\delta]} ({\vect{X}}) &= \limsup_{n \rightarrow \infty} \frac{1}{n} H_{[\delta]} (\TargetSr) . \label{eq:smooth_entropy} 
\end{align}
The following theorem indicates that the {$\mathrm{v}(\delta)$-resolvability} $S_{\rm v} (\delta|{\vect{X}})$ can be characterized by the smooth entropy for $\vect{X}$.
\begin{e_theo}[\cite{Yagi-Han2017}] \label{theo:d-VL_resolvability}
{\rm
For any general target source ${\vect{X}}$,
\begin{align}
S_{\rm v} (\delta|{\vect{X}}) = \lim_{\gamma \downarrow 0} H_{[\delta+\gamma]} ({\vect{X}}) ~~~(\delta \in [0, 1)). \label{eq:d-mean_resolvability_formula}
\end{align}}
\QED
\end{e_theo}

\section{Resolvability for Mixed Sources and Non-Mixed Channels} 

\subsection{Definitions}

In this section, the source $\vect{X} = \{ X^n \}_{n=1}^\infty $ is a \emph{mixed source} with general component sources.
Let $\Theta := \{ 1, 2 ,\cdots \} $ be the index set of component sources  $\vect{X}_i = \{X_i^n \}_{n=1}^\infty, i \in \Theta$, which may be a finite or countably infinite set.
The probability distribution of \emph{mixed source} $X^n$ is given by
\begin{align}
&P_{X^n}(\vect{x}) = \sum_{i \in \Theta } \alpha_i P_{X_i^n}(\vect{x})  ~~~(\forall n = 1,2,\cdots ; \forall \vect{x} \in \mathcal{X}^n), \label{eq:mixed_source1}
\end{align}
where $\alpha_i \ge 0$ with $\sum_{i \in \Theta} \alpha_i =1$.
Let $\vect{Y} = \{ Y^n \}_{n=1}^\infty$ be the channel output via $\vect{W}$ due to input $\vect{X}$. 
It is easily verified that the output distribution is given as a mixture of output distributions:
\begin{align}
&P_{Y^n}(\vect{y}) = \sum_{i \in \Theta } \alpha_i P_{Y_i^n}(\vect{y})~~~~~(\forall \vect{y} \in \mathcal{Y}^n), \label{eq:output_mixed_source1}
\end{align}
where $Y_i^n$ denotes the output via $W^n$ due to input $X_i^n$. 
The mixed source is formally denoted by\footnote{More generally, all results provided in this section hold for any mixed source with a \emph{general mixture}. Any stationary process can be characterized as a mixed source with {general mixture} whose components are ergodic processes.} $\{ (\vect{X}_i, \alpha_i) \}_{i \in \Theta} $.
Hereafter, the mixing ratio $\{\alpha_i\}_{i \in \Theta}$ is omitted if it is clear from the context, and we occasionally denote the mixed source simply by $\{\vect{X}_i\}$.

In this section, we consider a variant of the channel resolvability problems for mixed sources. 
Let $L_n^{(i)}$ denote a variable-length uniform random number for $i  \in \Theta$.
{Let} the random variable of length $L_n$ {be specified by}
\begin{align}
 \Pr\{  L_n = m \} =\sum_{i \in \Theta} \alpha_i \Pr\{ L_n^{(i)} = m\}  ~~(\forall m = 0, 1, 2, \cdots).
\end{align}
In other words, the length of a variable-length uniform random number $U^{(L_n)}$ obeys a mixture of the probability distributions for the lengths of component uniform random numbers $U^{(L_n^{(i)})}$. 
The average length of the uniform random number $U^{(L_n)}$ is given by
\begin{align}
\mathbb{E}[L_n] = \sum_{i \in \Theta} \alpha_i \mathbb{E}[L_n^{(i)}]. \label{eq:ave_length}
\end{align}
In the following problem, there are component encoders $\varphi_n^{(i)} : \mathcal{U}^* \rightarrow \mathcal{X}^n$, each of which approximates the channel output $Y_i^n$ via $W^n$ due to the $i$-th component source $X_i^n$.

\medskip
\begin{e_defin} \label{def:VL_achievable_rate3}
{\rm
Let $\delta \in [0, 1)$ be fixed arbitrarily.
A resolution rate $R \ge 0$ is said to be \emph{$\delta$-variable-length achievable} {or simply $\mathrm{v}(\delta)$-achievable} for mixed source $\{ ( \vect{X}_i, \alpha_i) \}_{i \in \Theta}$ {(under the variational distance)} if there exists a set of variable-length uniform random number $U^{(L_n^{(i)})}$ and a deterministic mapping $\varphi_n^{(i)} : \mathcal{U}^* \rightarrow \mathcal{X}^n$ satisfying
\begin{align}
 \limsup_{n \rightarrow \infty} \frac{1}{n} \E [L_n] &\le R, \label{eq:VL_ch_rate_cond3}  \\ 
 \limsup_{n \rightarrow \infty} \sum_{i \in \Theta } \alpha_i d(P_{Y_i^n}, P_{\tilde{Y}_i^n}) &\le \delta, \label{eq:VL_ch_variational_dist_cond3}   
\end{align}
where $\tilde{Y}_i^n $ denotes the output via $W^n$ due to the input $\tilde{X}_i^n = \varphi_n^{(i)}(U^{(L_n^{(i)})})$.
The infimum of all {$\mathrm{v}(\delta)$-achievable} rates for $\{ ( \vect{X}_i, \alpha_i) \}_{i \in \Theta}$:
\begin{align}
&{S_{\rm v}^\dagger} (\delta | \{\vect{X}_i \}, \vect{W}) \nonumber \\
&~~~:= \inf \{ R : ~ R ~\mbox{is {$\mathrm{v}(\delta)$-achievable} for}~ \{\vect{X}_i \}\} \label{eq:VL_ch_opt_rate3}
\end{align}
is called the \emph{$\delta$-variable-length channel resolvability} {or simply $\mathrm{v}(\delta)$-channel resolvability} for $\{ ( \vect{X}_i, \alpha_i) \}_{i \in \Theta}$.}
\QED
\end{e_defin}

\medskip
\begin{e_rema} \label{rema:rate_rel}
{\rm
In this problem, the condition for the approximation measure \eqref{eq:VL_ch_variational_dist_cond3} is changed from \eqref{eq:VL_ch_variational_dist_cond}.
It is well-known that the variational distance is \emph{jointly convex} in its arguments, and in general it holds that
\begin{align}
d(P_{Y^n}, P_{\tilde{Y}^n}) \le  \sum_{i} \alpha_i d(P_{Y_i^n}, P_{\tilde{Y}_i^n}),
\end{align}
where 
\begin{align}
P_{\tilde{Y}^n} (\vect{y}) = \sum_{i \in \Theta} \alpha_i  P_{\tilde{Y}_i^n} (\vect{y})~~~~(\forall \vect{y} \in \mathcal{Y}^n). 
\end{align}
Equation \eqref{eq:VL_ch_variational_dist_cond3}  imposes a more stringent condition than the one in \eqref{eq:VL_ch_variational_dist_cond}.
Since $S_{\rm v} (\delta | \vect{X} , \vect{W}) $ coincides with the $\delta$-\emph{mean} channel resolvability {\cite{Yagi-Han2017arXiv}}, for which the coin distribution may be any general source, in general we have 
\begin{align}
S_{\rm v} (\delta | \vect{X} , \vect{W}) \le {S_{\rm v}^\dagger} (\delta | \{\vect{X}_i \}, \vect{W}). \label{eq:rate_rel2}
 \end{align}} 
\QED
\end{e_rema}

\medskip
When the channel $W^n$ is an identity mapping, the addressed problem reduces to {that of} \emph{source resolvability} for $\{ \vect{X}_i \}$.
\medskip
\begin{e_defin} \label{def:VL_achievable_rate4}
{\rm
Assume that the channel  $W^n$ is an identity mapping.
The infimum of all {$\mathrm{v}(\delta)$-achievable} rates for $\{ ( \vect{X}_i, \alpha_i) \}_{i \in \Theta}$:
\begin{align}
{S_{\rm v}^\dagger} (\delta | \{ \vect{X}_i  \}):= \inf \{ R : ~ R ~\mbox{is {$\mathrm{v}(\delta)$-achievable} for}~ \{  \vect{X}_i \}\} \label{eq:VL_ch_opt_rate4}
\end{align}
is called the \emph{$\delta$-variable-length source resolvability} {or simply $\mathrm{v}(\delta)$-source resolvability} for $\{ ( \vect{X}_i, \alpha_i) \}_{i \in \Theta}$.}
\QED
\end{e_defin}

\subsection{Theorems}
\medskip
To characterize ${S_{\rm v}^\dagger} (\delta | \{\vect{X}_i\}, \vect{W} )$, we define 
 \begin{align}
  {H_{[\delta], W^n}^\dagger} (\{ X^n_i \}) := \inf_{\{P_{V_i^n}\} \in {B_\delta^\dagger}(\{X^n_i\}, W^n)} \sum_{i \in \Theta } \alpha_i H(V_i^n).
 \end{align} 
 where 
 \begin{align}
 &{B_\delta^\dagger} (\{X^n_i\}, W^n) \nonumber \\
  &~~= \Big\{  \{ P_{V_i^n}\}_{i \in \Theta} \subset \mathcal{P}(\mathcal{X}^n): \sum_{i \in \Theta} \alpha_i d(P_{Y_i^n}, P_{Z_i^n}) \le \delta \Big\},
\end{align}
where $Z_i^n$ denotes the output random variable via $W^n$ due to the input $V_i^n$.
In addition, we also define the asymptotic version:
\begin{align}
 {H_{[\delta], \vect{W}}^\dagger} (\{ \vect{X}_i \}) := \limsup_{n \rightarrow \infty} \frac{1}{n} {H_{[\delta], W^n}^\dagger} (\{  X^n_i \}).
\end{align}
Both ${H_{[\delta], W^n}^\dagger} (\{ X^n_i \})$ and ${H_{[\delta], \vect{W}}^\dagger} (\{ \vect{X}_i \})$ are nonincreasing {monotone} functions in $\delta$.
When the channel $\vect{W}$ is an identity mapping, ${H_{[\delta], W^n}^\dagger} (\{ X^n_i \})$ and ${H_{[\delta], \vect{W}}^\dagger} (\{ \vect{X}_i \})$ are denoted simply by ${H_{[\delta]}^\dagger} (\{ X^n_i \})$ and ${H_{[\delta]}^\dagger} (\{ \vect{X}_i \})$, respectively.
We establish the following theorem:
\begin{e_theo} \label{theo:mixed_channel_res_formula}
{\rm
For any mixed source $\vect{X} = \{ (\vect{X}_i, \alpha_i) \}_{i \in \Theta}$, it holds that
\begin{align}
{S_{\rm v}^\dagger} (\delta | \{ \vect{X}_i \}, \vect{W}) = \lim_{\gamma \downarrow 0} H_{[\delta+ \gamma], \vect{W}}^\dagger (\{ \vect{X}_i \})~~~~(\forall \delta \in [0,1)).
 \end{align}} 
\end{e_theo}
{(Proof)~~The proof is described in Sect.\ \ref{sect:proof_mixed_formula1}.}
\QED

When $\vect{W}$ is an identity mapping, we have the following corollary.
\begin{e_coro} \label{coro:mixed_source_res_formula}
{\rm
For any mixed source $\vect{X} = \{ (\vect{X}_i, \alpha_i) \}_{i \in \Theta}$, it holds that
\begin{align}
{S_{\rm v}^\dagger} (\delta | \{ \vect{X}_i \}) = \lim_{\gamma \downarrow 0} H_{[\delta+ \gamma]}^\dagger (\{ \vect{X}_i \})~~~~(\forall \delta \in [0,1)).
 \end{align}} 
\QED
\end{e_coro}
As is noted in Remark \ref{rema:rate_rel}, we have \eqref{eq:rate_rel2} in general. 
It is not clear if ${S_{\rm v}^\dagger} (\delta | \{ \vect{X}_i \}, \vect{W}) $ is equal to ${S_{\rm v}} (\delta | \vect{X}, \vect{W})$. 
The following theorem provides an interesting relationship between the two $\mathrm{v}(\delta)$-source resolvability problems for mixed sources. 
\begin{e_theo} \label{theo:source_res_equivalence}
{\rm
For any mixed source $\vect{X} = \{ ( \vect{X}_i, \alpha_i) \}_{i \in \Theta}$, it holds that
\begin{align}
S_{\rm v} (\delta | \vect{X}) = {S_{\rm v}^\dagger} (\delta | \{ \vect{X}_i\})~~~~(\forall \delta \in [0,1)). \label{eq:problem_eq}
 \end{align}} 
\end{e_theo}
{(Proof)~~The proof is described in Sect.\ \ref{sect:proof_mixed_formula2}.}
\QED

\begin{e_rema} \label{rema:FV_coding}
{\rm
The $\rm{v}(\delta)$-source resolvability $S_{\rm v} (\delta | \vect{X} )$ is equal to the minimum rate of the FV source coding achieving the decoding error probability asymptotically not greater than $\delta \in [0,1)$ \cite{Yagi-Han2017}.
We denote by $R_{\rm v}^*(\delta | \vect{X} )$ this minimum rate, and then from Theorem \ref{theo:source_res_equivalence}, we obtain
\begin{align}
R_{\rm v}^*(\delta | \vect{X} ) = S_{\rm v} (\delta | \vect{X}) = {S_{\rm v}^\dagger} (\delta | \{ \vect{X}_i\})~~~~(\forall \delta \in [0,1)).
\end{align}
for any mixed source  $\vect{X} = \{ ( \vect{X}_i, \alpha_i) \}_{i \in \Theta}$.
To characterize $R_{\rm v}^*(\delta | \vect{X} ) $, it suffices to analyze $S_{\rm v}^\dagger (\delta | \{ \vect{X}_i\})$, which may be easier for some mixed sources. In the succeeding sections, we demonstrate this claim for mixed memoryless sources. }
\QED
\end{e_rema}

%=======================================================
\section{Proof of Theorems \ref{theo:mixed_channel_res_formula} and \ref{theo:source_res_equivalence}}

\subsection{Proof of Theorem \ref{theo:mixed_channel_res_formula}} \label{sect:proof_mixed_formula1}

\emph{1) Converse Part:}~~~ Let $R$ be {$\mathrm{v}(\delta)$-achievable for $\{ \vect{X}_i\}$}. 
Then, there exists $U^{(L_n^{(i)})}$ and $\varphi_n^{(i)}$ satisfying \eqref{eq:VL_ch_rate_cond3}  and
\begin{align}
 &\limsup_{n \rightarrow \infty} \delta_n \le \delta, \label{eq:d-new_dist_cond}   
\end{align}
where we define 
\begin{align}
\delta_n = \sum_{i \in \Theta} \alpha_i d(P_{Y_i^n}, P_{\tilde{Y}_i^n})
\end{align}
and $\tilde{Y}_i^n$ is the output via $W^n$ due to the input $\tilde{X}_i^n= \varphi_n^{(i)} (U^{(L_n^{(i)})})$.
Equation \eqref{eq:d-new_dist_cond} implies that for any given $\gamma >0$, $\delta_n \le \delta + \gamma$ for all {$n \ge n_0$ with some $n_0 >0$}, and therefore
\begin{align}
H_{[\delta + \gamma], W^n}^\dagger (\{ X_i^n\} ) \le H_{[\delta_n], W^n}^\dagger(\{ X_i^n\} )~~~(\forall n \ge n_0) \label{eq:conv_ineq1d}
\end{align}
because ${H_{[\delta], W^n}^\dagger}(\{ X_i^n\} )$ is a nonincreasing {monotone} function of $\delta$.
Since $\{ P_{\tilde{X}_i^n} \} \subset B_{\delta_n}^\dagger (\{ X_i^n \}, W^n)$, we have
\begin{align}
H_{[\delta_n], W^n}^\dagger (\{ X_i^n\} ) \le \sum_{i \in \Theta} \alpha_i H(\tilde{X}_i^n). \label{eq:conv_ineq2d}
\end{align}
On the other hand, it follows that
\begin{align}
\sum_{i \in \Theta} \alpha_i H(\tilde{X}_i^n) & \le \sum_{i \in \Theta} \alpha_i H(U^{(L_n^{(i)})}) \nonumber \\
&= \sum_{i \in \Theta} \alpha_i \E [L_n^{(i)}] +  \sum_{i \in \Theta} \alpha_i H(L_n^{(i)}), \label{eq:conv_ineq3d}
\end{align}
where the inequality is due to the fact that $\varphi_n^{(i)}$ is a deterministic mapping and $\tilde{X}_i^n = \varphi_n^{(i)} (U^{(L_n^{(i)})})$.
By invoking the well-known relation (cf. \cite[Corollary 3.12]{Csiszar-Korner2011}) it holds that
\begin{align}
H(L_n^{(i)}) \le \log (e \cdot \E [L_n^{(i)}]). \label{eq:length_entropy_UB2}
\end{align}
In view of \eqref{eq:VL_ch_rate_cond3}, \eqref{eq:length_entropy_UB2} leads to
\begin{align}
&\limsup_{n \rightarrow \infty} \frac{1}{n} \sum_{i \in \Theta} \alpha_i  H(L_n^{(i)}) \nonumber \\
&~~~ \le \limsup_{n \rightarrow \infty} \frac{1}{n}  \sum_{i \in \Theta} \alpha_i  \log (e \cdot \E [L_n^{(i)}]) \nonumber \\
&~~~ \le\limsup_{n \rightarrow \infty}  \frac{1}{n}  \log \Big(e \cdot  \sum_{i \in \Theta} \alpha_i  \E [L_n^{(i)}] \Big)  = 0. \label{eq:vanishing_length_entropy}
\end{align}

Combining  \eqref{eq:conv_ineq1d}--\eqref{eq:conv_ineq3d} yields
\begin{align}
&H_{[\delta+\gamma], \vect{W}}^\dagger  (\{\vect{X}_i \}) \nonumber \\
&~~= \limsup_{n \rightarrow \infty} \frac{1}{n} H_{[\delta + \gamma], W^n}^\dagger  (\{X_i^n\}) \nonumber \\
&~~\le \limsup_{n \rightarrow \infty} \frac{1}{n} \E [L_n]  +  \limsup_{n \rightarrow \infty} \frac{1}{n} \sum_{i \in \Theta} \alpha_i H(L_n^{(i)}) \le R,  \nonumber
\end{align}
where we have used \eqref{eq:ave_length} for the first inequality and {\eqref{eq:VL_ch_rate_cond3} and} \eqref{eq:vanishing_length_entropy} for the second inequality.
Since $\gamma > 0$ is arbitrary, we obtain 
\begin{align}
\lim_{\gamma \downarrow 0} H_{[\delta+\gamma], \vect{W}}^\dagger  (\{\vect{X}_i \}) \le R.
\end{align}

\emph{2) Direct Part:}~~~ By the analogous argument to the proof of the direct part of Theorem \ref{theo:d-VL_resolvability} \cite{Yagi-Han2017}, we can show that the rate $R := H^* + 3 \gamma$ is $\mathrm{v}(\delta)$-achievable for $\{ \vect{X}_i \}$, where $H^*  = \lim_{\gamma' \downarrow 0} H_{[\delta+\gamma'], \vect{W}}^\dagger  (\{\vect{X}_i \})$ and $\gamma > 0$ is an {arbitrarily} small constant.
The proof sketch is as follows:

\begin{enumerate}

\item[(i)] We choose some $\{ P_{V_i^n} \}  \subset B_{\delta + \gamma}^\dagger  (\{ X_i^n\}, W^n)$ satisfying
\begin{align}
\sum_{i} \alpha_i H(V_i^n) \le H_{[\delta+\gamma], W^n}^\dagger  (\{X_i^n \}) + \gamma. 
\end{align}
By definition, we have 
\begin{align}
 \sum_{i \in \Theta} \alpha_i d(P_{Y_i^n}, P_{Z_i^n}) \le \delta + \gamma, \label{eq:dist_eval1}
\end{align}
where $Z_i^n$ denotes the output via $W^n$ due to the input $V_i^n$.

\item[(ii)]
Define
\begin{align}
S_n^{(i)}(m) := \left\{ \TargetSeq \in \CdAlphabet^n :  \left\lceil \log \frac{1}{P_{V_i^n}(\TargetSeq)} + n \gamma \right\rceil  = m\right\}.
\end{align}
For each $i \in \Theta$, we set
\begin{align}
\Pr[L_n^{(i)} = m] := \Pr[V_i^n \in S_n^{(i)}(m)].
\end{align}
In the same way as in the proof of Theorem \ref{theo:d-VL_resolvability} \cite{Yagi-Han2017}, we arrange {an} encoder $\varphi_n^{(i)}$ to generate $\tilde{X}_i^n = \varphi_n^{(i)}(U^{(L_n^{(i)})})$.

\item[(iii)] 
The average length rate can be evaluated as
\begin{align}
\mathbb{E}[L_n^{(i)}] \le  \left( 1 + \frac{1}{K^{n\gamma}} \right) \left( H(V_i^n) + n\gamma + 1 \right),  \label{eq:length_eval1}
\end{align}
whereas the variational distance satisfies
\begin{align}
d(P_{Z_i^n}, P_{\tilde{Y}_i^n}) &\le d(P_{V_i^n}, P_{\tilde{X}_i^n}) \le \frac{1}{2} K^{-n\gamma} + \gamma.  \label{eq:dist_eval2}
\end{align}

From \eqref{eq:length_eval1} and \eqref{eq:dist_eval2}, we obtain
\begin{align}
 \limsup_{n \rightarrow \infty} \frac{1}{n} \mathbb{E}[L_n]  & = \limsup_{n \rightarrow \infty} \frac{1}{n}  \sum_{i \in \Theta} \alpha_i \mathbb{E}[L_n^{(i)}] \nonumber \\
& \le  \limsup_{n \rightarrow \infty} \frac{1}{n} \sum_{i \in \Theta} \alpha_i H(V_i^n) + 2 \gamma \nonumber \\
&\le H^* + 3 \gamma = R
\end{align}
and
\begin{align}
 &\limsup_{n \rightarrow \infty} \sum_{i \in \Theta} \alpha_i d(P_{Y_i^n}, P_{\tilde{Y}_i^n}) \nonumber \\
 &~~\le \limsup_{n \rightarrow \infty}  \sum_{i \in \Theta} \alpha_i ( d(P_{Y_i^n}, P_{Z_i^n}) + d(P_{Z_i^n}, P_{\tilde{Y}_i^n}))  \nonumber \\
 &~~\le \limsup_{n \rightarrow \infty}  \sum_{i \in \Theta} \alpha_i d(P_{Y_i^n}, P_{Z_i^n}) + \gamma \le \delta + 2 \gamma,  \nonumber 
\end{align}
where the first inequality is due to the triangle inequality and the third inequality follows from \eqref{eq:dist_eval1}. Since $\gamma> 0$ is an arbitrary small constant, we conclude that $R$ is  $\mathrm{v}(\delta)$-achievable for $\{ \vect{X}_i \}$.
\end{enumerate}

\subsection{Proof of Theorem \ref{theo:source_res_equivalence}} \label{sect:proof_mixed_formula2}

Assume, without loss of generality, that the elements of $\mathcal{X}^n$ are indexed as $\vect{x}_1, \vect{x}_2, \cdots \in \mathcal{X}^n$ {so that}
\begin{align}
P_{X^n}(\vect{x}_j) \ge P_{X^n}(\vect{x}_{j+1})~~~(\forall j = 1, 2, \cdots).
\end{align}
For a given $\delta \in [0,1)$, let $j^*$ denote the integer satisfying
\begin{align}
\sum_{j=1}^{j^* - 1} P_{X^n}(\vect{x}_j) < 1 - \delta,~~~~~\sum_{j=1}^{j^*} P_{X^n}(\vect{x}_j) \ge 1 - \delta. \label{eq:j*}
\end{align}
Let $V_\delta^n$ be a random variable taking values in $\mathcal{X}^n$ whose probability distribution is given by
\begin{align}
P_{V_\delta^n} (\vect{x}_j) = \left\{ 
\begin{array}{ll}
P_{X^n}(\vect{x}_j) + \delta & \mathrm{for}~ j =1 \\
P_{X^n}(\vect{x}_j)  & \mathrm{for}~ j =2, 3, \cdots, j^* -1 \\
P_{X^n}(\vect{x}_j)  - \varepsilon &  \mathrm{for}~ j =j^*\\
 0 &  \mathrm{otherwise},
\end{array}
\right. \label{eq:min_dist}
\end{align}
where we define $\varepsilon = \delta - \sum_{j \ge j^*+1} P_{X^n}(\vect{x}_j)$.
It is easily checked that {$0 \le \varepsilon \le P_{X^n}(\vect{x}_{j^*})$ and} the probability distribution $P_{V_\delta^n}$ \emph{majorizes}\footnote{For a sequence $\vect{u}= (u_1, u_2, \cdots, u_L)$ of length $L$, we denote by $\tilde{\vect{u}} = (\tilde{u}_1, \tilde{u}_2, \cdots, \tilde{u}_L)$ the permuted version of $\vect{u}$ satisfying $\tilde{u}_i \ge \tilde{u}_{i+1}$ for all $i = 1, 2, \cdots, L$, where ties are arbitrarily broken.
We say $\vect{u}= (u_1, u_2, \cdots, u_L)$ \emph{majorizes} $\vect{v}= (v_1, v_2, \cdots, v_L)$ if $\sum_{i=1}^j \tilde{u}_i \ge \sum_{i=1}^j \tilde{v}_i$ for all $j=1, 2, \cdots, L$.} any $P_{V^n} \in B_\delta(X^n)$ \cite{Ho-Yeung2010}.
Since the Shannon entropy is a \emph{Schur concave} function\footnote{A function $f(\vect{u})$ is said to be \emph{Schur concave} if  $f(\vect{u}) \le f(\vect{v})$ for any pair $(\vect{u}, \vect{v})$ {such that} $\vect{v}$ is majorized by $\vect{u}$.}  \cite{MOA2011}, we immediately obtain the following lemma, which provides a characterization of $H_{[\delta]}(X^n)$.
\begin{e_lem}[\cite{Ho-Yeung2010}] \label{lem:smooth_entropy_compute}
\begin{align}
H_{[\delta]}(X^n) = H(V_\delta^n) ~~~(\forall \delta \in [0,1)).
\end{align}
\QED 
\end{e_lem}

{Let $j^*$ be the integer satisfying \eqref{eq:j*}.
Let $V^n$ be a random variable taking values in $\mathcal{X}^n$ whose probability distribution is given by
\begin{align}
P_{V^n} (\vect{x}_j) = \left\{ 
\begin{array}{ll}
P_{X^n}(\vect{x}_j)  & \mathrm{for}~ j =1, 2, \cdots, j^* -1 \\
\eta &  \mathrm{for}~ j =j^*\\
 0 &  \mathrm{otherwise},
\end{array}
\right.
\end{align}
where we define $\eta = \sum_{j \ge j^*} P_{X^n}(\vect{x}_j)$.}
To prove Theorem \ref{theo:source_res_equivalence}, the following lemma is of use.
\begin{e_lem} \label{lem:ex_mixed_source}
{\rm
Let $X^n = \{(X_i^n, \alpha_i)\}_{i \in \Theta}$ be a mixed source. Then,
\begin{align}
{H(V^n)} \le H_{[\delta]}(X^n) + \frac{2 \log e}{e} ~~~(\forall \delta \in [0,1)). \label{eq:H*_LB}
\end{align}}
\end{e_lem}
(\emph{Proof})~~
Let $V_\delta^n$ be defined as in \eqref{eq:min_dist}.
From Lemma \ref{lem:smooth_entropy_compute}, we have
\begin{align}
&H(V^n)  - H_{[\delta]} (X^n) \nonumber \\
&~~~=  H(V^n) - H(V_\delta^n) \nonumber \\
 &~~~ \le P_{V^n} (\vect{x}_1) \log \frac{1}{ P_{V^n} (\vect{x}_1)} +  P_{V^n} (\vect{x}_{j^*}) \log \frac{1}{ P_{V^n} (\vect{x}_{j^*})} \nonumber \\
 &~~~ \le \frac{2 \log e}{e}, \label{eq:H_relation1}
\end{align}
where the last inequality is due to $x \log x \ge - \frac{\log e}{e}$ for all $x > 0$.
\QED

For every $i \in \Theta$,  let $P_{V_i^n}$ be the probability distribution satisfying
\begin{align}
P_{V_i^n} (\vect{x}_j) = \left\{ 
\begin{array}{ll}
P_{X_i^n}(\vect{x}_j)  & \mathrm{for}~  j=1, 2, \cdots, j^* -1 \\
\eta_i &  \mathrm{for}~ j =j^*\\
 0 &  \mathrm{otherwise},
\end{array}
\right.
\end{align}
where we define $\eta_i = \sum_{j \ge j^*} P_{X_i^n}(\vect{x}_j)$.
Then, we can easily verify that
\begin{align}
P_{V^n} (\vect{x}) = \sum_{i \in \Theta} \alpha_i P_{V_i^n} (\vect{x}) ~~~~(\forall \vect{x} \in \mathcal{X}^n). \label{eq:mixed_V}
\end{align}
That is, $\{ (V_i^n, \alpha_i)\}_{i \in \Theta}$ is a mixed source.

Defining
\begin{align}
D_n^{(i)} = \Big\{  \vect{x} \in \mathcal{X}^n :  P_{V_i^n} (\vect{x}) >  P_{X_i^n} (\vect{x}) \Big\},
\end{align}
the average variational distance can be evaluated as
\begin{align}
\sum_{i \in \Theta} \alpha_i d(P_{X_i^n}, P_{V_i^n}) &= \sum_{i \in \Theta} \alpha_i \sum_{\vect{x} \in D_n^{(i)}}  ( P_{V_i^n} (\vect{x}) -  P_{X_i^n} (\vect{x})) \nonumber \\
&= \sum_{i \in \Theta} \alpha_i  ( P_{V_i^n} (\vect{x}_{j^*}) -  P_{X_i^n} (\vect{x}_{j^*})) \nonumber \\
&= \sum_{i \in \Theta} \alpha_i (\eta_i  -  P_{X_i^n} (\vect{x}_{j^*}) ) \nonumber \\ 
&= \sum_{i \in \Theta} \alpha_i \sum_{j > j^*} P_{X_i^n} (\vect{x}_j) \nonumber \\ 
& = \sum_{j > j^*} P_{X^n} (\vect{x}_j)  \le \delta, \label{eq:dist1}
\end{align}
where the inequality is due to \eqref{eq:j*}.
Since the Shannon entropy is a concave function, 
\eqref{eq:mixed_V} and \eqref{eq:dist1} imply that
\begin{align}
H(V^n) \ge \sum_{i \in \Theta} \alpha_i H(V_i^n) \ge {H_{[\delta]}^\dagger} (\{X_i^n\}). \label{eq:H_relation2}
\end{align}
Combining {\eqref{eq:H*_LB} and \eqref{eq:H_relation2} with Theorem \ref{theo:d-VL_resolvability} and Corollary \ref{coro:mixed_source_res_formula}}, we obtain
\begin{align}
S_{\rm v} (\delta | \vect{X} , \vect{W}) \ge {S_{\rm v}^\dagger} (\delta | \{\vect{X}_i \}, \vect{W}). \label{eq:rate_rel3}
 \end{align} 
The reverse inequality obviously holds (cf.\ \eqref{eq:rate_rel2}), and hence we obtain the claim.

\begin{e_rema}
{\rm
As is seen from the above proof arguments, Theorems \ref{theo:mixed_channel_res_formula} and \ref{theo:source_res_equivalence} hold even with general probability space $\Theta$.}
\QED
\end{e_rema}

%=======================================================
\section{Resolvability for Mixed Memoryless Sources} 

In this section, we assume that the source $\vect{X} = \{ X^n \}_{n=1}^\infty $ is a \emph{mixed memoryless source} and the channel $\vect{W}$ is an identity mapping.
Each component source $\vect{X}_i = \{X_i^n \}_{n=1}^\infty, i  \in \Theta$ is stationary and memoryless, which is specified by a  source $X_i$ {over} $\mathcal{X}$ as
\begin{align}
&P_{X_i^n}(\vect{x}) = \prod_{j = 1}^n P_{X_i} (x_j) ~~~(\forall \vect{x}=(x_1, x_2, \ldots, x_n) \in \mathcal{X}^n). \label{eq:mixed_memoryless_source}
\end{align}
Without loss of essential generality, we assume that 
\begin{align}
+ \infty> H (X_1) \ge H (X_2) \ge \cdots, \label{eq:component_order}
\end{align}
where the component sources $\{X_i\}_{i \in \Theta}$ are indexed in the decreasing order of $H(X_i)$.

For given $\delta \in [0,1)$, we define the positive integer $i^*$ satisfying 
\begin{align}
\sum_{i < i^*} \alpha_i \le \delta,~~~~~A_{i^*}:=\sum_{i \le i^*} \alpha_i > \delta.  \label{eq:i^*}
\end{align}
We demonstrate an application of the general relationship \eqref{eq:problem_eq} between the two variable-length resolvability problems {to establish} a single-letter formula for the $\mathrm{v}(\delta)$-source resolvability.
\begin{e_theo}\label{theo:mixed_VL_ch_resolvability}
{\rm
For any  mixed memoryless source $\vect{X} = \{ (\vect{X}_i, \alpha_i) \}_{i \in \Theta}$, it holds that 
\begin{align}
S_{\rm v} (\delta | \vect{X} ) &= {S_{\rm v}^\dagger} (\delta | \{\vect{X}_i\} ) \nonumber \\
&=
 (A_{i^*}-\delta) H (X_{i^*}) + \sum_{i > i^*} \alpha_i  H(X_i).\label{eq:d-mixed_VL_source_formula}
\end{align}
for all $\delta \in [0,1)$.}
\QED
\end{e_theo}

\begin{e_rema}
{\rm
As was mentioned in Remark \ref{rema:FV_coding}, we have  $S_{\rm v} (\delta | \vect{X} ) = R_{\rm v}^*(\delta | \vect{X} )$ for all $\delta \in [0, 1)$ for any general source $\vect{X}$, where $R_{\rm v}^*(\delta | \vect{X} )$ denotes the minimum rate of the FV source coding achieving the decoding error probability asymptotically not greater than $\delta \in [0,1)$.
For mixed \emph{memoryless} source $\vect{X} = \{ \vect{X}_i \}$, Koga and Yamamoto \cite{Koga-Yamamoto2005} (for $\Theta$ with $|\Theta| = 2$) and  Kuzuoka \cite{Kuzuoka2016} (for any finite $\Theta$) have shown that $R_{\rm v}^*(\delta | \vect{X} )$ is characterized as
\begin{align}
R_{\rm v}^* (\delta | \vect{X} ) = (A_{i^*}-\delta) H (X_{i^*}) + \sum_{i > i^*} \alpha_i  H(X_i)  \label{eq:d-mixed_source_coding_formula}
\end{align}
for all $\delta \in [0,1)$ if the source alphabet $\mathcal{X}$ is \emph{finite}.
Since formula \eqref{eq:d-mixed_VL_source_formula} holds for any countably infinite $\Theta$ and $\mathcal{X}$, the relation $S_{\rm v} (\delta | \vect{X} ) = R_{\rm v}^*(\delta | \vect{X} )$ implies that formula \eqref{eq:d-mixed_source_coding_formula} actually holds for a wider class of mixed memoryless sources. 

\QED
}
\end{e_rema}

Since any stationary memoryless source is a mixed source with a singleton set $\Theta$, we immediately obtain the following corollary.
\begin{e_coro}[\cite{Koga-Yamamoto2005,Yagi-Han2017}]
{\rm
Let $\vect{X}$ be a stationary memoryless source $X$. Then, it holds that  
 \begin{align}
S_{\rm v} (\delta | \vect{X})  = (1-\delta) H (X) \label{eq:d-memoryless_VL_source_formula}
\end{align}
for all $\delta \in [0,1)$.}
\QED
\end{e_coro}

\medskip
\noindent
(\emph{Proof of Theorem \ref{theo:mixed_VL_ch_resolvability}})

The following argument demonstrates the usefulness of the general relationship \eqref{eq:problem_eq} between the two variable-length resolvability problems. 
Since {it holds that}
\begin{align}
S_{\rm v} (\delta | \vect{X}) =  S_{\rm v}^\dagger (\delta | \{\vect{X}_i\} ) = \lim_{\gamma \downarrow 0} H_{[\delta + \gamma]}^\dagger  (\{ \vect{X}_i\})
\end{align}
as is shown in Corollary \ref{coro:mixed_source_res_formula}, we first focus on the quantity $H_{[\delta ]}^\dagger  (\{X_i^n\})$.
The $\delta$-ball $B_{\delta}^\dagger  (\{X_i^n\})$, which is defined as $B_{\delta}^\dagger  (\{X_i^n\}, W^n)$ with an identity mapping $W^n$, can be written as
\begin{align}
&\hspace*{-3mm}  B_{\delta}^\dagger  (\{X_i^n\}) \nonumber \\
 &= \{ \{ P_{V_i^n}\} \subset \mathcal{P}(\mathcal{X}^n): \exists \{\delta_i \ge 0\} ~\mathrm{s.t.}~\sum_{i} \alpha_i \delta_i = \delta, \nonumber \\
 &~~~~~~~~~~~~~~~~~~~~~~~~~~~~~~~d(P_{X_i^n}, P_{V_i^n})\le \delta_i , \forall i \in \Theta\} \nonumber \\
&= \bigcup_{\{ \delta_i \ge 0:  \, \sum_i \alpha_i \delta_i = \delta \}} \bigcup_{i \in \Theta} B_{\delta_i} (X_i^n).
\end{align}
Then, it obviously holds that
\begin{align}
& \hspace*{-3mm} {H_{[\delta]}^\dagger} (\{ X_i^n \}) \nonumber \\
&= \inf_{\{ P_{V_i^n}\} \in {B_\delta^\dagger} (\{X_i^n \}) } \sum_{i \in \Theta} \alpha_i H(V_i^n) \nonumber \\
&=  \inf_{\{ \delta_i \ge 0: \, \sum_i \alpha_i \delta_i = \delta \}}   \inf_{P_{V_i^n} \in B_{\delta_i} (X_i^n) } \sum_{i \in \Theta} \alpha_i H(V_i^n) \label{eq:H_express} \\
&\ge \inf_{\{ \delta_i \ge 0: \, \sum_i \alpha_i \delta_i = \delta \}}   \sum_{i \in \Theta} \alpha_i  \inf_{P_{V_i^n} \in B_{\delta_i} (X_i^n) } H(V_i^n) \nonumber \\
&= \inf_{\{ \delta_i \ge 0: \, \sum_i \alpha_i \delta_i = \delta \}}   \sum_{i \in \Theta} \alpha_i H_{[\delta_i]} ( X_i^n ). \label{eq:H^dag_LB}
\end{align}
It is known (cf.\ \cite{Yagi-Nomura2017}) that 
\begin{align}
\liminf_{n \rightarrow \infty} \frac{1}{n} H_{[\delta]} ( X_i^n ) = (1 - \delta) H(X_i)  ~~~(\forall \delta \in [0,1))
\end{align}
for any stationary memoryless source $\vect{X}_i = \{X_i^n\}_{n = 1}^\infty$, and thus
\begin{align}
{H_{[\delta]}^\dagger} (\{ \vect{X}_i \})  &= \limsup_{n \rightarrow \infty} \frac{1}{n} {H_{[\delta]}^\dagger} (\{ X_i^n \}) \nonumber \\
&\ge \liminf_{n \rightarrow \infty} \frac{1}{n} {H_{[\delta]}^\dagger} (\{ X_i^n \})
\nonumber \\
&\ge \inf_{\{ \delta_i \ge 0: \, \sum_i \alpha_i \delta_i = \delta \}}   \sum_{i \in \Theta} \alpha_i \liminf_{n \rightarrow \infty} \frac{1}{n}   H_{[\delta_i]} ( X_i^n )
\nonumber \\\
&= \inf_{\{ \delta_i \ge 0: \, \sum_i \alpha_i \delta_i = \delta \}}   \sum_{i \in \Theta} \alpha_i  (1 - \delta_i) H(X_i) \nonumber \\
&= \inf_{\{ \alpha_i \ge \varepsilon_i \ge 0: \, \sum_i \varepsilon_i = \delta \}}   \sum_{i \in \Theta} ( \alpha_i - \varepsilon_i)  H(X_i), \label{eq:H_LB2}
\end{align}
where the second inequality is due to Fatou's lemma. Noticing that {the $\inf$ in} \eqref{eq:H_LB2} is a linear program and in view of \eqref{eq:component_order}, we find that the solution is given by 
\begin{align}
\varepsilon_i = \left\{
\begin{array}{ll}
\alpha_i & \mathrm{for} ~ i < i^* \\
\delta - \sum_{i < i^*} \alpha_i & \mathrm{for} ~ i = i^* \\
0 & \mathrm{for} ~ i > i^*,
\end{array}
\right.
\end{align}
yielding
\begin{align}
{H_{[\delta]}^\dagger} (\{ \vect{X}_i \}) \ge  (A_{i^*}-\delta) H (X_{i^*}) + \sum_{i > i^*} \alpha_i  H(X_i).
 \label{eq:H_LB3}
\end{align}
The right-hand side is right-continuous in $\delta \ge 0$, and thus it follows from Corollary \ref{coro:mixed_source_res_formula} that
\begin{align}
S_{\rm v} (\delta | \vect{X} ) &= {S_{\rm v}^\dagger} (\delta | \{\vect{X}_i\} ) \nonumber \\
& \ge  (A_{i^*}-\delta) H (X_{i^*}) + \sum_{i > i^*} \alpha_i  H(X_i),\label{eq:d-mixed_VL_source_LB}
\end{align}
{where it should be noted that $i^* = i^*(\delta)$ is right-continuous in $\delta$.}

To show the reverse inequality, we start {with} the characterization \eqref{eq:H_express}. We choose
\begin{align}
\delta_i = \left\{
\begin{array}{ll}
1 & \mathrm{for} ~ i < i^* \\
\frac{ \delta - \sum_{i < i^*} \alpha_i }{\alpha_i} & \mathrm{for} ~ i = i^* \\
0 & \mathrm{for} ~ i > i^*.
\end{array}
\right. \label{eq:delta_i}
\end{align}
We also set probability distributions $\{P_{V_i^n}\}$ on $\mathcal{X}^n$ by
\begin{align}
P_{V_i^n} (\vect{x}) = \left\{
\begin{array}{ll}
\Delta (\vect{x}) & \mathrm{for} ~ i < i^* \\
(1- \delta_i) P_{X_i^n} (\vect{x}) + \delta_i \Delta(\vect{x}) & \mathrm{for} ~ i = i^* \\
P_{X_i^n}(\vect{x})  & \mathrm{for} ~ i > i^*,
\end{array}
\right. \label{eq:prob_setting2}
\end{align}
where $\Delta(\vect{x}) = \vect{1} \{ \vect{x} = \vect{x}_0 \} $ is the delta distribution with some specific $\vect{x}_0 \in \mathcal{X}^n$. 
Then, it is easily verified that
\begin{align}
& \delta_i \ge 0~(\forall i \in \Theta) ~~\mathrm{s.t.}~~ \sum_{i \in \Theta} \alpha_i \delta_i = \delta,   \\
&d(P_{X_i^n}, P_{V_i^n}) \le \delta_i ~~~(\forall i \in \Theta), \label{eq:constraint_meet}
\end{align}
meaning $P_{V_i^n} \in B_{\delta_i} (X_i^n)$ for all $i \in \Theta$. 
Also, $H(V_{i^*}^n)$ can be evaluated as
\begin{align}
&\hspace*{-1mm}  H(V_{i^*}^n) \nonumber \\
& = \sum_{\vect{x} \in \mathcal{X}^n\setminus\{\vect{x}_0\}} P_{V_{i^*}^n} (\vect{x}) \log \frac{1}{(1- \delta_{i^*}) P_{X_{i^*}^n} (\vect{x})}  \nonumber \\
&~~+ P_{V_{i^*}^n} (\vect{x}_0) \log \frac{1}{P_{V_{i^*}^n} (\vect{x}_0)} \nonumber \\
&\le  \sum_{\vect{x} \in \mathcal{X}^n}  (1- \delta_{i^*}) P_{X_{i^*}^n} (\vect{x}) \log \frac{1}{(1- \delta_{i^*}) P_{X_{i^*}^n} (\vect{x})} + \frac{\log e}{e} \nonumber \\
&\le (1- \delta_{i^*}) H(X_{i^*}^n) + \frac{2 \log e}{e}, \label{eq:H_entropy_UB}
\end{align}
where the inequalities are due to $x \log x \ge - \frac{\log e}{e}$ for all $x \ge 0$.
With these choices of $\{ \delta_i\} $ and $\{P_{V_i^n}\}$ satisfying \eqref{eq:prob_setting2}--\eqref{eq:constraint_meet}, it follows from \eqref{eq:H_express} that
\begin{align}
\frac{1}{n} {H_{[\delta]}^\dagger} (\{ X_i^n \}) & \le \frac{1}{n}  \sum_{i \in \Theta} \alpha_i H(V_i^n) \nonumber \\
&= \frac{1}{n}  \sum_{i \ge i^*} \alpha_i  H(V_i^n) \nonumber \\
&=  \frac{\alpha_{i^*}}{n}  H(V_{i^*}^n) + \sum_{i > i^*}  \alpha_i H(X_i) .
\end{align}
Taking the limit superior in $n$ on both sides, we obtain 
\begin{align}
&\limsup_{n \rightarrow \infty} \frac{1}{n} {H_{[\delta]}^\dagger} (\{ X_i^n \}) \nonumber \\
&~~ \le \limsup_{n \rightarrow \infty} \frac{\alpha_{i^*}}{n}  H(V_{i^*}^n) + \sum_{i > i^*}  \alpha_i H(X_i)  \nonumber \\.
 &~~ \le\alpha_{i^*}(1 - \delta_{i^*})  H(X_{i^*}) + \sum_{i > i^*}  \alpha_i H(X_i)  \nonumber \\
 &~~ =  (A_{i^*}-\delta) H (X_{i^*}) + \sum_{i > i^*} \alpha_i  H(X_i),
\end{align}
where the second inequality follows from \eqref{eq:H_entropy_UB}.
Again, the right-hand side is right-continuous in $\delta \ge 0$, and Corollary \ref{coro:mixed_source_res_formula} indicates that 
\begin{align}
S_{\rm v} (\delta | \vect{X} ) &= {S_{\rm v}^\dagger} (\delta | \{\vect{X}_i\} ) \nonumber \\
& \le  (A_{i^*}-\delta) H (X_{i^*}) + \sum_{i > i^*} \alpha_i  H(X_i).\label{eq:d-mixed_VL_source_UB}
\end{align}
We complete the proof. 
\QED

%=======================================================
\section{Second-Order Resolvability for Mixed Sources} 

\subsection{Definitions} 

In this section, we generalize the addressed problems to the second order case.
The first definition corresponds to Definition \ref{def:VL_achievable_rate} in the first order \cite{Yagi-Han2017}.
\begin{e_defin}
{\rm
A second-order rate $L \in  (-\infty, + \infty)$ is said to be {\emph{$\mathrm{v}(\delta, R)$-achievable}} {(under {the} variational distance)} for $\vect{X}$ with $\delta \in [0, 1)$ and $R \ge 0$ if there exists a variable-length uniform random number $U^{(L_n)}$ and
a deterministic mapping $\varphi_n : \SrAlphabet^* \rightarrow \CdAlphabet^n$ satisfying
\begin{align}
 \limsup_{n \rightarrow \infty} \frac{1}{\sqrt{n}} \left( \E [L_n] - nR \right) &\le L, \label{eq:dR-variable_length_rate_cond}  \\ 
 \limsup_{n \rightarrow \infty} d(P_{X^n}, P_{\tilde{X}^n}) &\le \delta, \label{eq:dR-variable_length_variational_dist_cond}   
\end{align}
where $\ApproxSr = \varphi_n(U^{(L_n)})$ {and $\mathbb{E}[L_n]$ is specified as in \eqref{eq:ave_length}}.
The infimum of all $\mathrm{v}(\delta, R)$-achievable rates for $\vect{X}$ is denoted by
\begin{align}
T_{\rm v} (\delta, R| {\vect{X}}) &:= \inf\{ L : ~ L ~\mbox{is $\mathrm{v}(\delta, R)$-achievable~for}~\vect{X} \}. \label{eq:dR-variable_length_opt_rate}
\end{align}}
\QED
\end{e_defin}

We also consider a variant problem for mixed sources $\vect{X} = \{ \vect{X}_i \}$.

\begin{e_defin} \label{def:2nd_VL_achievable_rate1}
{\rm
A second-order rate $L \in  (-\infty, + \infty)$ is said to be {\emph{$\mathrm{v}(\delta, R)$-achievable}} {(under {the} variational distance)} for mixed source $\{ ( \vect{X}_i, \alpha_i) \}_{i \in \Theta}$ {with $\delta \in [0, 1)$ and $R \ge 0$} if there exists a set of variable-length uniform random number $U^{(L_n^{(i)})}$ and a deterministic mapping $\varphi_n^{(i)} : \mathcal{U}^* \rightarrow \mathcal{X}^n$ satisfying
\begin{align}
 \limsup_{n \rightarrow \infty} \frac{1}{\sqrt{n}} \left( \E [L_n] - nR \right) &\le R, \label{eq:2nd_VL_ch_rate_cond2}  \\ 
 \limsup_{n \rightarrow \infty} \sum_{i \in \Theta } \alpha_i d(P_{X_i^n}, P_{\tilde{X}_i^n}) &\le \delta, \label{eq:VL_ch_variational_dist_cond4}   
\end{align}
where $\tilde{X}_i^n = \varphi_n^{(i)}(U^{(L_n^{(i)})})$.
The infimum of all $\mathrm{v}(\delta, R)$-achievable rates for $\{ ( \vect{X}_i, \alpha_i) \}_{i \in \Theta}$ is denote by:
\begin{align}
T_{\rm v}^\dagger (\delta, R | \{\vect{X}_i \}) &:= \inf \{ L : ~L ~\mbox{is~$\mathrm{v}(\delta, R)$-achievable} \nonumber \\
&~~~~~~~~~~~~~~~~~ \mbox{for}~\{ \vect{X}_i\} \}. \label{eq:VL_source_opt_rate3}
\end{align}}
\QED
\end{e_defin}

\begin{e_rema} \label{rema:interesting_second_order}
{\rm
It is easily verified that 
\begin{align}
T_{\rm v} (\delta, R|{\vect{X}})  = \left\{
\begin{array}{ll}
+  \infty & \mbox{for}~ R < S_{\rm v}(\delta|{\vect{X}})  \\
-  \infty & \mbox{for}~ R > S_{\rm v}(\delta|{\vect{X}}).
\end{array}
\right.
\end{align}
Hence, only the case $R = S_{\rm v}(\delta|{\vect{X}})$ is of our interest.
The same remark also applies to $T_{\rm v}^\dagger (\delta, R| \{ \vect{X}_i \})$}.
\QED
\end{e_rema}

\subsection{Theorems} 

The following theorems indicate that {$T_{\rm v} (\delta, R|{\vect{X}})$ and $T_{\rm v}^\dagger (\delta, R|\{\vect{X}_i\})$} can also be characterized by the smooth entropies.
\begin{e_theo} \label{theo:2nd_d-VL_resolvability}
{\rm
For any mixed source $\{( \vect{X}_i, \alpha_i) \}_{i \in \Theta}$,
\begin{align}
\hspace*{-2mm} T_{\rm v} (\delta, R|{\vect{X}}) &= \lim_{\gamma \downarrow 0} \limsup_{n \rightarrow \infty} \frac{1}{\sqrt{n}} ( H_{[\delta+\gamma]} (X^n) - nR), \label{eq:2nd_formula1} \\
\hspace*{-2mm} T_{\rm v}^\dagger (\delta, R|\{ \vect{X}_i\}) &= \lim_{\gamma \downarrow 0} \limsup_{n \rightarrow \infty} \! \frac{1}{\sqrt{n}}  ( H_{[\delta+\gamma]}^\dagger (\{X_i^n\}) \! - \! nR) \label{eq:2nd_formula2}
\end{align}
for all $\delta \in [0, 1)$ and $R \ge 0$.} 
\end{e_theo}
(\emph{Proof})~~For the proof of \eqref{eq:2nd_formula1}, see \cite{Yagi-Han2017}.
Formula \eqref{eq:2nd_formula2} can be proven in a parallel way to Theorem \ref{theo:mixed_channel_res_formula}.
\QED 

\medskip
As in the first order case, we have the equivalence between $T_{\rm v} (\delta, R|{\vect{X}}) $ and $T_{\rm v}^\dagger (\delta, R|\{ \vect{X}_i\}) $ for any mixed source $\{ \vect{X}_i \}$.
\begin{e_theo} \label{theo:2nd_d-VL_resolvability3}
{\rm
For any mixed source $\{( \vect{X}_i, \alpha_i) \}_{i \in \Theta}$,
\begin{align}
\hspace*{-2mm} T_{\rm v} (\delta, R|{\vect{X}}) = T_{\rm v}^\dagger (\delta, R|\{ \vect{X}_i\})~~~(\delta \in [0, 1), R \ge 0). \label{eq:d-mean_resolvability_formula2}
\end{align}} 
\end{e_theo}
(\emph{Proof})~~This theorem can be proven in a parallel way to Theorem \ref{theo:source_res_equivalence}.
\QED 

\medskip
We now turn to analyzing the $\mathrm{v}(\delta,R)$-source resolvability for mixed \emph{memoryless} sources.
We assume the following properties:
\begin{enumerate}
\item[(i)] The index set $\Theta$ is finite. 

\item[(ii)] Each component source $X_i$ has the finite third absolute moment of $\log\frac{1}{P_{X_i}(X_i)}$.

\item[(iii)] {Component sources $\{X_i \}$ satisfy 
\begin{align}
+ \infty > H(X_1) > H(X_2) > \cdots.
\end{align}}
\end{enumerate}
The following lemma is useful to establish a single-letter formula of the $\mathrm{v}(\delta, R)$-source resolvability.
\begin{e_lem}[\cite{KPV2015}] \label{lem:KPV}
{\rm
Assume that a stationary memoryless source $X^n$ has a finite absolute moment of $\log\frac{1}{P_{X}(X)}$. Then, it holds that 
\begin{align}
H_{[\delta]}(X^n) = (1 - \delta) n H(X)&  - \textstyle \sqrt{\frac{n V(X)}{2 \pi}} e^{-\frac{(Q^{-1}(\delta))^2}{2}} \nonumber \\
 &+ O(1),
\end{align}
where  $V(X)$ denotes the variance of $\log \frac{1}{P_{X}(X)}$ (varentropy) and $Q^{-1}$ denotes the inverse of the complementary cumulative distribution function of the standard Gaussian distribution.}
 \QED
\end{e_lem}

\begin{e_theo} \label{theo:2nd_d-VL_resolvability4}
{\rm
Let $\vect{X} = \{( \vect{X}_i, \alpha_i) \}_{i \in \Theta}$ be a mixed memoryless source satisfying (i)--(iii). For {$R = S_{\rm v}(\delta|\vect{X}) = S_{\rm v}^\dagger (\delta|\{\vect{X}_i\})$} given by \eqref{eq:d-mixed_VL_source_formula}, it holds that
\begin{align}
\hspace*{-2mm} T_{\rm v} (\delta, R|{\vect{X}}) &= T_{\rm v}^\dagger (\delta, R|\{ \vect{X}_i\}) \nonumber \\
&= -  {\alpha_{i^*} \textstyle  \sqrt{\frac{V(X_{i^*})}{2 \pi}} e^{-\frac{(Q^{-1}(\delta_{i^*}))^2}{2}}}, \label{eq:d-mean_resolvability_formula3}
\end{align}
where $i^*$ is the integer satisfying \eqref{eq:i^*} and {$\delta_{i^*}$} is defined as in \eqref{eq:delta_i}.} 
\end{e_theo}
(\emph{Proof})~~The direct part is comparatively easy and we omit the proof due to the space limitation.

 To prove the converse part, we define $D(\delta) :=\{ \{ \delta_i \}: \delta_i \ge 0, \sum_{i \in \Theta} \alpha_i \delta_i = \delta \}$. 
Using \eqref{eq:H^dag_LB} and Lemma \ref{lem:KPV}, we obtain
 \begin{align}
\hspace*{-3mm} \frac{1}{\sqrt{n}} H_{[\delta]}^\dagger (\{X_i^n\}) &\ge \! \inf_{\{\tilde{\delta}_i \} \in D(\delta)} \sum_{i \in \Theta} \alpha_i \Big\{ (1 - \tilde{\delta_i}) \sqrt{n} H(X_i) \nonumber \\
 &~~~~ - \textstyle \sqrt{\frac{ V(X_i)}{2 \pi}} e^{-\frac{(Q^{-1}(\tilde{\delta}_i))^2}{2}} + o(1) \Big\},
 \end{align}
 and thus for all $n > n_0$ with some $n_0>0$ the minimizer $\{\tilde{\delta}_i\} \in D(\delta)$ on the right-hand side is $\{\delta_i\}$ given in \eqref{eq:delta_i}: i.e.,
  \begin{align}
\hspace*{-3mm} \frac{1}{\sqrt{n}} H_{[\delta]}^\dagger(\{X_i^n\}) &\ge \sum_{i \ge i^*} \alpha_i (1 - \delta_i) \sqrt{n} H(X_i) \nonumber \\
 &~~ - \textstyle  {\alpha_{i^*} \sqrt{\frac{ V(X_{i^*})}{2 \pi}} e^{-\frac{(Q^{-1}(\delta_{i^*}))^2}{2}}}  + o(1)
 \end{align}
 for all $n > n_0$, {where we have used the fact that $e^{-\frac{(Q^{-1}(0))^2}{2}} \allowbreak = 0$.} 
 Since $R = S_{\rm v}(\delta | \vect{X})$ is given by
 \begin{align}
 R = \sum_{i \ge i^*} \alpha_i (1-\delta_i) H(X_i)
 \end{align}
 due to Theorem \ref{theo:mixed_VL_ch_resolvability}, it follows that
 \begin{align}
&\lim_{\gamma \downarrow 0} \limsup_{n \rightarrow \infty} \frac{1}{\sqrt{n}} (H_{[\delta + \gamma]}^\dagger(\{X_i^n\}) - nR) \nonumber \\
 &~~\ge - {\alpha_{i^*} \textstyle \sqrt{\frac{ V(X_{i^*})}{2 \pi}} e^{-\frac{(Q^{-1}(\delta_{i^*}))^2}{2}}}. 
 \end{align}
In view of Theorems \ref{theo:2nd_d-VL_resolvability} and \ref{theo:2nd_d-VL_resolvability3}, we complete the proof of the converse part.
\QED 

\medskip
\begin{e_rema}
{\rm
As in the first order case, $T_{\rm v}(\delta, R| \vect{X})$ is equal to $R_{\rm v}^* (\delta, R | \vect{X})$, which denotes the minimum achievable rate of the FV $\delta$-source coding \cite{Yagi-Nomura2017}. 
Theorem \ref{theo:2nd_d-VL_resolvability4} also indicates that
\begin{align}
\hspace*{-2mm} R_{\rm v}^* (\delta, R | \vect{X}) = -  {\alpha_{i^*} \textstyle \sqrt{\frac{ V(X_{i^*})}{2 \pi}} e^{-\frac{(Q^{-1}(\delta_{i^*}))^2}{2}}}, \label{eq:d-mean_resolvability_formula4}
\end{align}
for a mixed memoryless source $\vect{X} = \{ (\vect{X}_i, \alpha_i) \}_{i \in \Theta}$ satisfying (i)--(iii).}
\end{e_rema}


\begin{thebibliography}{10}
{\small
\bibitem{Csiszar-Korner2011}
I.~Csisz{\'a}r and J.~K{\"{o}}rner, \emph{Information Theory: Coding Theorems for Discrete Memoryless Systems}, 2nd ed., Cambridge University Press, Cambridge, U.K., 2011.

\bibitem{Han2003}
T.~S.\ Han, \emph{Information-Spectrum Methods in Information Theory}, Springer, 2003.

\bibitem{Han-Verdu93}
T.~S.\ Han and S.\ Verd\'u, ``Approximation theory of output statistics,'' \emph{IEEE Trans. Inf. Theory}, vol.~39, no.~3, pp.~752--771, May 1993.

\bibitem{Hayashi2006}
M.~Hayashi, ``General nonasymptotic and asymptotic formulas in channel resolvability and identification capacity and their application to the wiretap channel,'' \emph{IEEE Trans.\ Inf.\ Theory}, vol.\ 52, no.\ 4, Apr. 2006.

\bibitem{Ho-Yeung2010}
S.-W.\ Ho and R.~W.\ Yeung, ``The interplay between entropy and variational distance,'' \emph{IEEE Trans. Inf. Theory}, vol.~56, no.~12, pp.~5906--5929, Dec.\ 2010.

\bibitem{Koga-Yamamoto2005}
H.\ Koga and H.\ Yamamoto, ``Asymptotic properties on codeword length of an optimal FV code for general sources,'' \emph{IEEE Trans.\ Inf.\ Theory}, vol.\ 51, no.\ 4, pp.\ 1546--1555, Apr. 2005.

\bibitem{KPV2015}
V.~Kostina, Y.~Polyanskiy, and S.~Verd\'{u}, ``Variable-length compression allowing errors,'' \emph{IEEE Trans. Inf. Theory}, vol.~61, no.~8, pp. 4316--4330, Aug.\ 2015.

\bibitem{Kuzuoka2016}
S.\ Kuzuoka, ``Variable-length coding for mixed sources with side information allowing decoding errors,'' \emph{Proc. Int. Symp. on Inform. Theory and its Applications}, Oct.\ 2016.

\bibitem{MOA2011}
A.\ W.\ Marshall, I.\ Olkin, and B.C.\ Arnold, \emph{Inequalities: Theory of Majorization and Its Applications}, 2nd Ed.\, Springer, New York, NY, 2011.
\bibitem{Yagi2017}
H.~Yagi, ``Characterizations of fixed-length resolvability for general sources and channels,'' \emph{Proc.\ 2017 IEEE Int. Symp. on Inf. Theory}, Jun.\ 2017.

\bibitem{Yagi-Han2017arXiv}
H.~Yagi and T.\ S.~Han, ``Variable-length resolvability for general sources and channels,'' \emph{arXiv:1701.08712}, Jan.\ 2017.

\bibitem{Yagi-Han2017}
H.~Yagi and T.\ S.~Han, ``Variable-length resolvability for general sources,'' \emph{Proc.\ 2017 IEEE Int. Symp. on Inf. Theory}, Jun.\ 2017.

\bibitem{Yagi-Han2017sita}
H.~Yagi and T.\ S.~Han, ``Variable-length resolvability for discrete memoryless channels,'' submitted to \emph{Proc.\ 2018 IEEE Int. Symp. on Inf. Theory}, Jun.\ 2018.

\bibitem{Yagi-Nomura2017}
H.~Yagi and R.~Nomura, ``Variable-length coding with cost allowing non-vanishing error probability,'' \emph{IEICE Trans. Fundamentals}, vol.\ E100-A, no.\ 8, pp.\ 1683--1692, Aug.\ 2017.

} 
\end{thebibliography}
\end{document}